# 3C 345: the historical light curve (1966-1991) from the digitized plates of the Asiago Observatory

Alessandro Omizzolo[1], Cesare Barbieri[2], Corinne Rossi[3],

[1] Vatican Observatory, Castelgandolfo, [2] Department of Astronomy, University of Padova, [3] University of Roma I,

## Abstract

In the frame of a large project to digitize the plate archives of the Italian and Vatican Astronomical Observatory, we have already performed the digitization of all available plates of the field of 3C345. The plates, approximately 100, were taken with the three telescopes of the Asiago Observatory (122 cm, 180 cm, 67/90 Schmidt Telescope) in the period from 1966 to 1991. Many of them were not analyzed till now, so we present here essentially new data about the variability of 3C 345 and also of other six quasars in the same field, in that period.

## Introduction

Many plates were taken at the Asiago Observatory of selected quasars, to study their optical variability (see for instance in Barbieri,1977). Well followed was in particular 3C 345, one of the most luminous and violently variable ones. Part of these data have been published by Barbieri et al. (1977) but many exposures were taken after that publication. Here we present the light curve covering the entire period of the Asiago observations; while the paper of 1977 gave the magnitudes estimated by traditional means on the original photographic emulsion, here we present the results of the photometric analysis of the digitized plates. This work is part of a larger project aimed to preserve the plate archives of several Italian Observatories and of the Vatican Observatory (see Barbieri et al., 2003). The systematic scanning activity of the plates of these observatories was initiated by giving priority to the more densely covered fields (such as the field of M31, M33, 3C 345, etc.) for specific scientific goals, in this case long-term QSO optical variability. As detailed in Barbieri et al. (2003), a good quality commercial scanner with a resolution of 1600 dpi is used, acquiring the images at 14 bit by using a dedicated software whose output is a fits file.

## The field of 3C 345

The plates of the 3C 345 field were taken mainly in the B band (103aO+GG13, or IIaO+GG13), with few plates also in V (103aD+GG11, 103aD+GG14) and U (103aO+UG2). In Table 1 we report the census of the plates used for this research.

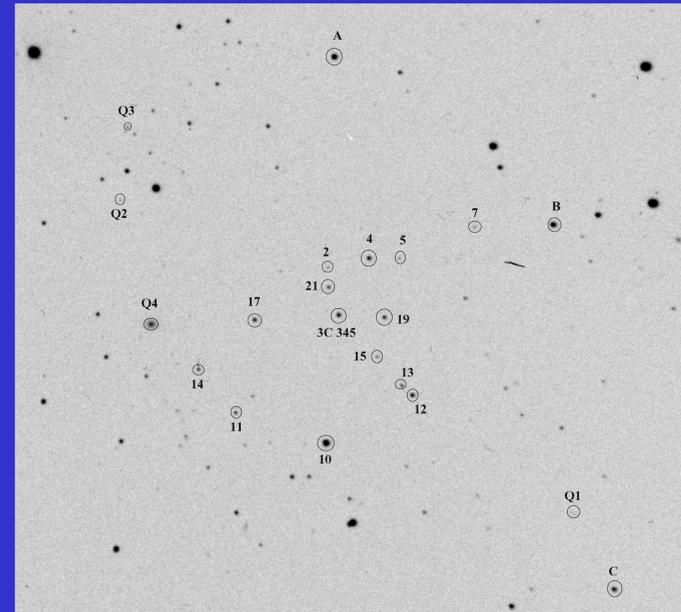

Figure 1 –3C 345 with the quasars (Q1-Q4) in its field (plate 8031) and the photometric sequence used.

| Telescope/Band | U | B | V |
|---|---|---|---|
| Schmidt 67/90 | | 90 | |
| Newtonian 122 | 5 | 29 | 5 |
| Cassegrain 180 | | 11 | 1 |
| Total | 5 | 90 | 6 |

Table 1 – The plates used to build the light curve

| Name | α2000 | δ2000 | z |
|---|---|---|---|
| Q1 | 16 42 28.5 | +39 43 43 | 1.86 |
| Q2 | 16 43 27.01 | +39 51 25 | 1.083 |
| Q3 | 16 43 26.2 | +39 53 14 | 0.704 |
| Q4 | 16 43 23.1 | +39 48 24 | 0.30 |
| Q5 | 16 43 19.5 | +39 43 39 | 2.0 |
| Q6 | 16 43 34.9 | +39 53 47 | 0.594 |

Table 2 – The positions and redshifts of the quasars in the field

We limited the reduction of the images to the field common to all three telescopes (namely, the field of the 180 cm telescope), where not only 3C345 but also other 6 QSOs are present in the catalog of Veron (2003), as reported in Table 2 :

In Fig. 1 we report the field of 3C 345 with the photometric sequence used in this work and the position of the other QSOs. In a future paper we'll present the results for the much larger number of QSOs present on the S67/92 images, which cover approximately 25 sq degrees. We also plan to study in detail the periodicity in the light curve to confirm the conclusions published by Barbieri at al. (1977), Zhang et al. (1998)., Zhang et al. (2000), Caproni et al. (2003).

## The photometric reduction

The main part of the work to obtain the magnitude of the QSO consisted in setting up a standard procedure for the calibration of the scanned images, in order to use standard software packages (i.e. IRAF, in particular the tasks DAOPHOT and those relate to it). The calibration has been derived using the magnitudes published by Angione (1971) and by Gonzales-Perez et al. (2001), as reported in Table 3. Three stars are in common. Owing to the larger number and magnitude coverage we have used the sequence of Gonzales-Perez et al. The agreement between the IRAF magnitudes and the photometric sequence is as detailed in Table 4. The three telescopes provide magnitudes having essentially the same error.

## The light Curve

A preliminary light curve in the B band is reported in fig.2. With respect to Barbieri et al. (1997) we present new data from 1977 to 1990. This new light curve is in good agreement with the general one obtained by Zhang (1998) and it confirms the rather complicated behavior of the luminosity variations. The overall behavior can be described as a continuous activity at a level of ± 0.3 mag (see for example the variations in correspondence to about MJD 40500) that increase by a factor of roughly two (see the variations in correspondence to about MJD 41000) when there is an outburst. We report also the unpublished data in the V and U bands (see Table 1). The color index B-V is about +0.2; the data are few but they are coherent with the behavior reported by Zhang (2000) relative to more recent observations obtained with CCD detector.
For the other 6 quasars the variability range is of about 1.5 for Q1, about 2 mag for Q2, and 1 mag for Q3. Q4 is a galaxy known as NGC 6212, and we can surely detect strong variability of its nucleus at the level of 2 mag (see fig. 3 for the light curve of these quasars). For Q5 and Q6 we have only two data! A fuller study of these other objects will be the subject of a future paper.

| Star | U Gonzales | B Gonzales | V Gonzales | U Angione | B Angione | V Angione |
|---|---|---|---|---|---|---|
| 1 | | 19.22 | 17.52 | | | |
| 2 | 18.26 | 18.46 | 17.86 | | | |
| 3 | | 19.03 | 17.88 | | | |
| 4, D | 16.43 | 16.04 | 15.25 | 16.73 | 16.08 | 15.18 |
| 5 | | 17.92 | 16.71 | | | |
| 6 | | 17.27 | 18.27 | | | |
| 7 | | 17.78 | 17.12 | | | |
| 8 | | 19.11 | 18.24 | | | |
| 9 | | 16.37 | 15.74 | | | |
| 10 | 13.89 | 13.96 | 13.77 | | | |
| 11 | | 17.04 | 16.44 | | | |
| 12 | 16.14 | 16.05 | 15.42 | | | |
| 13 | 16.97 | 17.24 | 16.72 | | | |
| 14 | | 17.09 | 16.08 | | | |
| 15 | 17.52 | 17.64 | 17.13 | | | |
| 16 | | 19.22 | 18.17 | | | |
| 17 | | 16.34 | 15.7 | | | |
| 18 | | 18.66 | 16.9 | | | |
| 19, F | 16.93 | 16.45 | 15.23 | 17.58 | 16.47 | 15.22 |
| 21, E | 16.32 | 17.08 | 16.51 | | 17.17 | 16.52 |
| A | | | | 14.34 | 14.44 | 13.76 |
| B | | | | 14.71 | 14.85 | 14.14 |
| C | | | | 15.83 | 15.87 | 15.29 |

Tab. 3 – Synopsis of the two photometric sequences for 3C 345

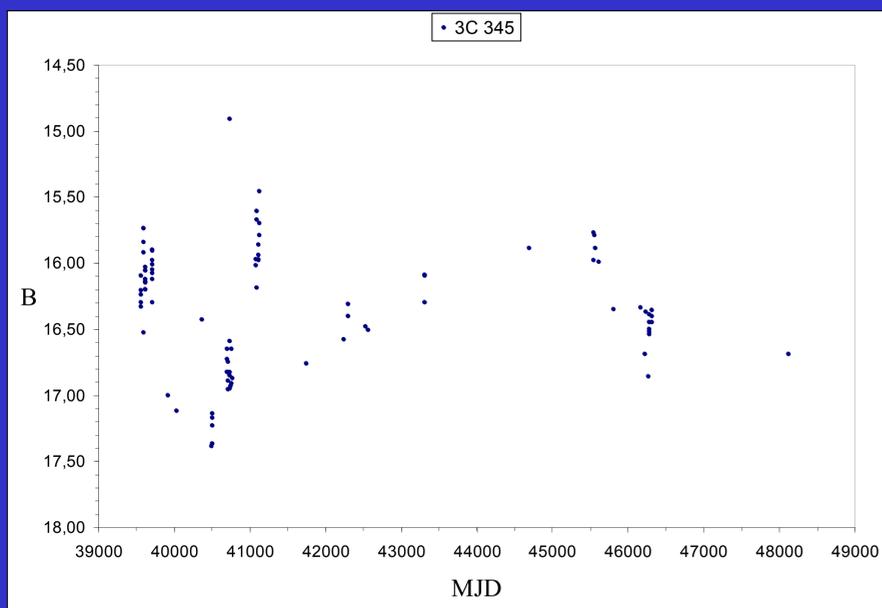

Figure 2 –The light curve of 3C 345 in the B band..

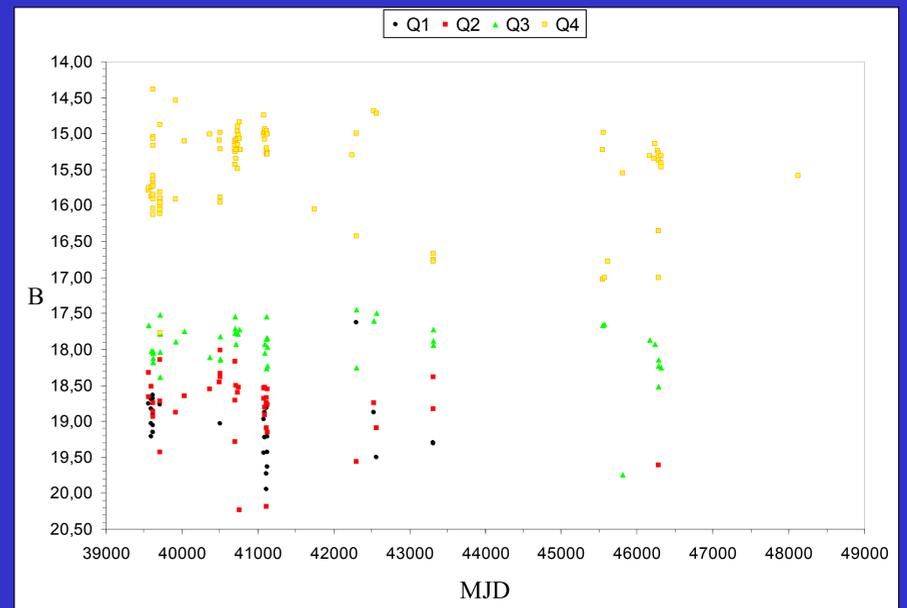

Figure 3 –The light curve of Q1, Q2, Q3, Q4 in the B band..

| Telescope | σ(B) (mag) | Standard deviation of σ(B) |
|---|---|---|
| S 67/92 | 0,14 | 0,03 |
| Newtonian 122 | 0,17 | 0,05 |
| Cassegrain 180 | 0,15 | 0,05 |

Tab. 4– Standard deviation of the B magnitudes

Acknowledgment: We thank S. Magrin and F. Rampazzi for the kind help in the digitization and analysis of the data. This work has been performed in the the frame of a National project financed by the Italian Ministry for University and Research (MIUR) in 2002.